\documentclass[parskip=full, titlepage=false, twoside=false]{scrartcl} 
\usepackage[utf8]{inputenc} 
\usepackage[T1]{fontenc} 
\usepackage[english]{babel}
\usepackage[autostyle=true]{csquotes}
\usepackage{lmodern}
\usepackage[pdfauthor={Jonathan Glöckle}, pdftitle={Initial data rigidity via Dirac-Witten operators}]{hyperref} 
\usepackage{amsmath, amsthm, amssymb, bm, bbm, mathrsfs, mathtools}
\usepackage{extarrows}
\usepackage{tabulary,tabularx}
\usepackage{graphicx}
\usepackage{tikz-cd}
\usepackage{enumerate}
\usepackage[capitalise]{cleveref}
\usepackage{bbold}

\theoremstyle{plain} 
\newtheorem{Satz}{Theorem} 
\newtheorem{Prop}[Satz]{Proposition} 
\newtheorem{Lem}[Satz]{Lemma}
 
\theoremstyle{definition}

\newtheorem{Bem}[Satz]{Remark}

\newtheorem{Frag}[Satz]{Question}

\crefname{Satz}{Theorem}{Theorems}
\crefname{Prop}{Proposition}{Propositions}
\crefname{Lem}{Lemma}{Lemmata}
\crefname{Kor}{Corollary}{Corollaries}
\crefname{Def}{Definition}{Definitions}

\newcommand{\arxiv}[1]{\textsc{arxiv}: \href{https://arxiv.org/abs/#1}{\nolinkurl{#1}}}

\let\div\relax

\newcommand{\Z}{\mathbbm{Z}} 
 
\newcommand{\R}{\mathbbm{R}}

\newcommand{\del}{\partial}

\newcommand{\upd}{\mathrm{d}}

\renewcommand{\emptyset}{\varnothing}
\renewcommand{\epsilon}{\varepsilon}
\renewcommand{\phi}{\varphi}
\renewcommand{\hat}{\widehat}

\DeclareMathOperator{\div}{div}

\DeclareMathOperator{\tr}{tr}

\DeclareMathOperator{\scal}{scal}
\DeclareMathOperator{\ric}{ric}

\DeclareMathOperator{\ein}{ein}

\begin{document}
\author{Jonathan Glöckle\thanks{Fakultät für Mathematik, Universität Regensburg, 93040 Regensburg, Germany, E-mail address: \href{mailto:jonathan.gloeckle@mathematik.uni-regensburg.de}{jonathan.gloeckle@mathematik.uni-regensburg.de}}}
\title{Initial data sets with dominant energy condition admitting no smooth dec spacetime extension}
\date{\today}
\maketitle

\begin{abstract}
	There are two versions of the dominant energy condition (=dec):
	The original one for Lorentzian manifolds and an associated one for initial data sets.
	If a Lorentzian manifold satisfies dec, then so does the induced initial set on any embedded spacelike hypersurface.
	In this article, we discuss the question of a potential converse of this:
	Is every dec initial data set the induced one on a spacelike hypersurface within a suitably chosen dec Lorentzian manifold?
	We provide an example showing that in general the answer is no if we require all structures to be smooth.	
\end{abstract}

Relativity theory describes the universe as an $n+1$-dimensional Lorentzian manifold $(\overline{M}, \overline{g})$, which may be equipped with fields encoding the matter distribution.
It is supposed to satisfy the Einstein field equation $\ein = T$, where $\ein = \ric^{\overline{g}} - \frac{1}{2} \scal^{\overline{g}} \overline{g}$ is the Einstein curvature and $T$ is (up to a potential constant) the energy-momentum tensor associated to the matter data.
Whereas for a specific matter model an explicit expression of $T$ in terms of the matter field may be written down, doing so seems to be an impossible task for the universe we live in.
For this reason cosmological considerations rely on general physical properties $T$ should have rather than on its explicit form.
These properties are known as energy conditions.
In this article, we focus on one of those:
The (spacetime) \emph{dominant energy condition} (=dec) requires that $T(V,W) \geq 0$ for all causal vectors $V,W \in T_p\overline{M}$ lying in the same connected component of the light cone in $T_p\overline{M}$, $p \in \overline{M}$.
Note that via the Einstein equation, it becomes a curvature condition for $(\overline{M}, \overline{g})$.

In the following, Lorentzian manifolds $(\overline{M}, \overline{g})$ will usually be time-oriented and we use the term \emph{spacetime} for such. 
We now consider an embedded spacelike hypersurface $M \subset \overline{M}$ of a spacetime.
This carries an induced initial data set $(g,k)$, where $g$ is the induced metric on $M$ and $k = g(\nabla^{\overline{g}}_- e_0, -)$ is the second fundamental form with respect to the future unit normal $e_0$ of $M$ in $\overline{M}$.
In general, by an \emph{initial data set} on a manifold~$M$ we mean a pair $(g,k)$ consisting of a Riemannian metric $g$ and a symmetric $2$-tensor field $k$ on $M$.
\emph{Energy density} and \emph{momentum density} of $(g,k)$, respectively, are defined by
\begin{align*}
	\rho &= \frac{1}{2}(\scal^g + \tr^g(k) - |k|_g^2) \\
	j &= \div^g k - \upd \tr^g(k).
\end{align*}
By the Gauß-Codazzi equations, for the induced initial data set on a spacelike hypersurface we have $\ein(e_0, -) = -\rho \overline{g}(e_0, -) + j$.
Hence, if $(\overline{M}, \overline{g})$ satisfies the dec, then  $\rho \geq |j|_g$.
Motivated by this, for an initial data set $(g,k)$ the condition $\rho \geq |j|_g$ is known as (initial data) \emph{dominant energy condition}.

The initial data dec is a main assumption in the (spacetime) positive mass theorem, whose first instances were proved back in the 80s \cite{Schoen.Yau:1981,Witten:1981,Parker.Taubes:1982} but which lately lead to new results applying to non-spin manifolds \cite{Eichmair.Huang.Lee.Schoen:2016}, involving boundaries \cite{Galloway.Lee:2021} and focusing on the equality case \cite{Huang.Lee:2020,Huang.Lee:2023p}.
A more detailed (and more complete) account of these developments can be found in the recent textbook by Dan Lee \cite{Lee:2019}.
Other results involving the initial data dec concern Hawking's black hole topology theorem \cite{Galloway.Schoen:2006}, topological censorship \cite{Andersson.Dahl.Galloway.Pollack:2018} as well as the Penrose inequality and related inequalities \cite{Cha.Khuri.Sakovich:2016}.
Localized deformations of dec initial data on a fixed manifold were studied by Corvino and Huang \cite{Corvino.Huang:2020} and non-connectedness results for the space of initial data with strict dec were obtained by the author \cite{Gloeckle:2023,Gloeckle:2021p}.

It is easy to see that every initial data set $(g,k)$ on any manifold $M$ arises as induced initial data set on a spacelike hypersurface within a spacetime.
In fact, $M = M \times \{0\} \subset (\overline{M} = M \times \R, \overline{g} = g_t - \upd t^2)$ does the job when $(g_t)_{t \in \R}$ is a smooth family of Riemannian metrics on $M$ with $g_0 = g$ and $\frac{\upd}{\upd t}_{|t = 0} g_t = 2 k$.
It seems natural to ask whether such a statement is also true when dec is required:
\begin{Frag} \label{Que:Main}
	Given an initial data set $(g,k)$ on a manifold $M$ satisfying $\rho \geq |j|_g$, does there exist a \emph{dec} spacetime $(\overline{M}, \overline{g})$ containing $M$ as spacelike hypersurface with induced initial data set $(g,k)$?
\end{Frag}
Phrased more conceptually, \cref{Que:Main} asks whether dec initial data sets provide sufficient initial conditions to solve a certain partial differential inequality (called ``Einstein inequality'' by Rendall \cite{Rendall:1992}).
In \cite{Ammann.Gloeckle:2023} Bernd Ammann and the author conjectured that the answer to \cref{Que:Main} is yes.
This conjecture is supported by two special cases:
Firstly, if $\rho = 0$ and $j = 0$, then a solution of the vacuum Cauchy problem in general relativity yields an extension to a spacetime with $\ein = 0$ and thus subject to dec.
In fact, for any dec spacetime extension the domain of dependence of the spacelike hypersurface has to be a vacuum solution due to the conservation theorem from \cite[Sec.~4.3]{Hawking.Ellis:1973}.
Secondly, the conjecture also holds when dec is satisfied in the strict sense $\rho > |j|_g$.
\begin{Prop}[{\cite[Prop.~1.10]{Gloeckle:2019}}] \label{Prop:Sdec}
	If an initial data set $(g,k)$ on a manifold $M$ satisfies $\rho > |j|_g$, then there is a dec spacetime $(\overline{M}, \overline{g})$ containing $M$ as spacelike hypersurface with induced initial data set $(g,k)$.
\end{Prop}

The proof of \cref{Prop:Sdec} roughly consists of two steps.
First, finding for $(\rho, j)$ a symmetric $2$-tensor $\hat{T} \in \Gamma(T^*M \odot T^*M)$ -- the \emph{stress} part of the energy-momentum tensor -- such that
\begin{gather*} 
	T(xe_0 + X, ye_0 + Y) \coloneqq \rho xy + xj(Y) + yj(X) + \hat{T}(X,Y) > 0
\end{gather*}
whenever $x,y > 0$, $X,Y \in T_pM$ satisfy $|X|_g \leq x$, $|Y|_g \leq y$.
Second, constructing a spacetime $(\overline{M}, \overline{g})$ that contains $M$ as spacelike hypersurface such that the induced initial data set is $(g,k)$ and $\ein_{|TM \otimes TM} = \hat{T}$.
Then, by construction, for every $p \in M$ the strict condition that $\ein(V,W) > 0$ for all future-causal vectors $V,W \in T_p\overline{M}$ is satisfied.
Since this strict version of dec is an open condition, $(\overline{M}, \overline{g})$ satisfies dec -- after possibly restricting $\overline{M}$ to a small open neighborhood of $M$.

Somewhat surprisingly, already the analog of the first step cannot be carried out in the general case, where dec may also hold non-strictly.
More precisely, let $(g,k)$ be an initial data set subject to $\rho \geq |j|_g$.
It is not hard to find for every $p \in M$ a $\hat{T}_p \in T_p^*M \odot T_p^*M$ with
\begin{gather} \label{eq:dec}
		\rho_p xy + xj_p(Y) + yj_p(X) + \hat{T}_p(X,Y) \geq 0 \quad\text{when } |X|_g \leq x,\, |Y|_g \leq y,
\end{gather}
where $x,y \in \R$ and $X,Y \in T_pM$.
In fact, 
\begin{gather} \label{eq:DefT}
	\hat{T}_p = 
	\begin{cases}
	\frac{j_p \otimes j_p}{\rho_p} & \rho_p \neq 0 \\
	0 & \rho_p = 0
	\end{cases}
\end{gather}
works.
However, the section of $T^*M \odot T^*M \to M$ defined by \eqref{eq:DefT} is in general only of regularity $C^1$ (\cref{Rem:C1Reg}) but, as we shall see, not $C^2$.
A priori, of course, there might be other choices of $p \mapsto \hat{T}_p$ that are smooth in addition to satisfying \eqref{eq:dec}.
But if $\rho = |j|_g$, there is no flexibility:
\begin{Lem} \label{Lem:Rigid}
	Suppose $T_pM$ is a real vector space with scalar product $g_p$, $\rho_p \in \R$ and $j_p \in (T_pM)^*$ with $\rho_p = |j_p|_{g_p}$.
	Then $\hat{T}_p \in (T_pM)^* \otimes (T_pM)^*$ satisfies \eqref{eq:dec} if and only if it is given by \eqref{eq:DefT}.
\end{Lem}

Using these observations we construct an example showing that in general \cref{Que:Main} has to be answered negatively in the smooth setting.
\begin{Satz}[Main Theorem] \label{Thm:Main}
	For every manifold $M$ of dimension $n \geq 3$, there is a (smooth) dec initial data set $(g,k)$ such that on a non-empty open subset $U \subset M$ the following hold:
	\begin{itemize}
		\item $\rho = |j|_g$ and
		\item $p \mapsto \hat{T}_p$ defined by \eqref{eq:DefT} is not $C^2$. 
	\end{itemize}
	Any such initial data set does not arise as induced initial data set on a spacelike hypersurface within a \emph{smooth} dec spacetime.
\end{Satz}

Finally, it might be interesting to note the following.
Initial data sets $(g,k)$ as in \cref{Thm:Main} cannot be completed to initial data for the Einstein-Vlasov system (cf.~\cite[eq.~(7.13)-(7.15)]{Ringstroem:2013}), although this matter model -- describing an ensemble of massive particles that are only gravitationally interacting -- is very flexible.
The reason is that a solution spacetime $(\overline{M}, \overline{g})$ for the Einstein-Vlasov system (which exists) would both satisfy the dominant energy condition and contain $M$ with induced initial data set $(g,k)$.
The same reasoning also applies to any other matter model (which is supposed to satisfy dec) with well-posed Cauchy problem.
Of course, also the condition $\rho = |j|_g$ is far from what is expected from massive particles, and thus for matter of Vlasov type.

\paragraph*{Acknowledgements}
I want to thank Bernd Ammann for pointing out to me the relevance of the question discussed here and for his continuous support.
Thanks to Carla Cederbaum for her comments on the difference between the dominant energy conditions.
I also thank Dan Lee, Hans Ringström, Mathias Dahl and Olaf Müller for some initial discussions on how to possibly find an affirmative answer to \cref{Que:Main}.
Parts of this article were written during a research stay at the ESI Vienna.
I received financial support from the SFB~1085 funded by the Deutsche Forschungsgemeinschaft.

\section*{Proofs of the results}
This section contains the proofs for the claims made in the introduction.
We start off by showing the rigidity statement for $\hat{T}_p$ if $\rho_p = |j_p|_{g}$. 
\begin{proof}[Proof of \Cref{Lem:Rigid}]
	To ease notation in the proof we leave out the subscript $p$ for $\rho_p$, $j_p$, $\hat{T}_p$ and $g_p$. 
	Let us first note that the tensor $\hat{T}$ from \eqref{eq:DefT} satisfies \eqref{eq:dec}, which is trivial in the case $\rho = 0$ and in the case $\rho \neq 0$ follows from
	\begin{gather*}
		\rho xy + xj(Y) + yj(X) + \hat{T}(X,Y) = \frac{1}{\rho}(\rho x + j(X))(\rho y + j(Y))
	\end{gather*}
	along with $|j(X)| \leq |j|_g|X|_g \leq \rho x$ and $|j(Y)| \leq |j|_g|Y|_g \leq \rho y$ for all $x,y \in \R$, $X,Y \in T_pM$ with $|X|_g \leq x$, $|Y|_g \leq y$.
	
	The converse is also easy to see if $\rho = 0$, since the condition $\hat{T}(X,Y) \geq 0$ for all $X,Y \in T_pM$ implies $\hat{T} = 0$.
	In the case $\rho \neq 0$ fix $X \colon (-\epsilon, \epsilon) \to T_pM$ with $X(0) = -j^\sharp$ and $|X(t)|_g = \rho \eqqcolon x$ for all $t \in (-\epsilon, \epsilon)$.
	Then consider the function
	\begin{gather*}
		F(t,Y) \coloneqq \rho^2 |Y|_g + \rho j(Y) + |Y|_g j(X(t)) + \hat{T}(X(t),Y)
	\end{gather*}
	for $t \in (-\epsilon, \epsilon)$ and $Y \in T_pM$.
	The dominant energy condition~\eqref{eq:dec} implies $F(t,Y) \geq 0$ for all $t \in (-\epsilon,\epsilon)$ and $Y \in T_pM$.
	Since $j(j^\sharp) = |j|_g^2 = \rho^2$, the function reduces to $F(0,Y) = \rho j(Y) - \hat{T}(j^\sharp, Y)$ for $t=0$.
	Replacing $Y$ by $-Y$, we see that non-negativity of $F$ implies $F(0,-) \equiv 0$ or, equivalently, $\hat{T}(j^\sharp,-) = \rho j$.
	Furthermore, we must have $\frac{\upd}{\upd t}_{|t=0} F(t,Y) = 0$ for all $Y \in T_pM$, which yields $\hat{T}(X^\prime(0), Y) = 0$ (since $t \mapsto j(X(t))$ attains a minimum in $t=0$).
	
	Now, given any $X,Y \in T_pM$, we split up $X = \frac{j(X)}{|j|_g^2}j^\sharp + X_\perp$.
	Since $X_\perp \perp j^\sharp$, we can choose a curve $X \colon (-\epsilon,\epsilon) \to T_pM$ as above with $X^\prime(0) = X_\perp$.
	We get
	\begin{align*}
		\hat{T}(X,Y) &= \frac{j(X)}{|j|_g^2}\hat{T}(j^\sharp, Y) + \hat{T}(X^\prime(0),Y) \\
			&= \frac{j(X)}{|j|_g^2} \rho j(Y) + 0 = \frac{j(X)j(Y)}{\rho}. \qedhere
	\end{align*}
\end{proof}

From now on, we consider the function $\hat{T} \colon p \to \hat{T}_p$ defined by \eqref{eq:DefT} and study its regularity.

\begin{Bem} \label{Rem:C1Reg}
	It is easy to see from $|j|_g \leq \rho$ that $\hat{T}$ is continuous.
	In fact, it is even in $C^1$ as we will show now.
	Clearly it is smooth wherever $\rho \neq 0$, so we only have to consider points $p \in M$ with $\rho_p = 0$.
	Let us first consider the case $\upd_p \rho \neq 0$.
	In this case, $\rho = 0$ defines a smooth hypersurface locally around $p$ and after choosing an appropriate chart of $M$, we may assume without loss of generality that $M \subset \R^n$ is an open subset, $p = 0$ and $\rho$ vanishes exactly on $M \cap (\{0\} \times \R^{n-1})$.
	After possibly restricting further to a relatively compact open neighborhood of $0$, there exists some $C > 0$ such that the components of $j$ satisfy $|j_i| \leq C\rho$ for $i =1, \ldots, n$.
	In particular, each $j_i$ also vanishes along the hypersurface defined by $x_1 = 0$.
	In general, given a smooth function $f \colon (-\epsilon,\epsilon)^n \to \R$ with $f(0,x_2, \ldots, x_n) = 0$ for all $x_2, \ldots, x_n$, the function $\tilde{f} = \frac{f}{x_1}$ is also smooth:
	Firstly, we may assume without loss of generality that $f$ is compactly supported.
	Then the Lemma of Borel (cf.~\cite[Thm.~1.2.6]{Hoermander:2003_2nd}) is applicable, allowing us to find a smooth function $g$ such that for all $k \geq 0$
	\begin{gather*}
		\frac{1}{k!}\frac{\del^k}{\del x_1^k} g = \frac{1}{(k+1)!}\frac{\del^{k+1}}{\del x_1^{k+1}} f
	\end{gather*}
	holds on the hypersurface $x_1=0$.
	Secondly, since $f-x_1 g$ and all its (higher) derivatives in normal direction vanish along $x_1 = 0$, all partial derivatives of $h = \frac{f- x_1 g}{x_1}$ continuously extend by zero along $x_1 = 0$.
	Thus $\tilde{f} = g + h$ is indeed a smooth function.
	Applying this to $\rho$ and $j_i$ and noting $\tilde{\rho}(0) = \lim_{x \to 0} \frac{\rho}{x_1} = \frac{\del \rho}{\del x_1} (0) \neq 0$, we see that
	\begin{align*}
		\frac{j_i}{\rho} &= \frac{x_1 \tilde{j_i}}{x_1 \tilde{\rho}} = \frac{\tilde{j}_i}{\tilde{\rho}}
	\end{align*}
	uniquely extends to a smooth function in a neighborhood of $0$.
	Hence, near $p$, we may consider $\frac{j}{\rho}$ as a smooth $1$-form and so $\hat{T} = \frac{j}{\rho} \otimes j$ is also smooth there.
	
	Let us look at points with $\upd_p \rho = 0$.
	Note first that in any such point and for any $X \in T_pM$ we have $\nabla_X j = 0$. The reason is that for a smooth curve $\gamma \colon (-\epsilon, \epsilon) \to M$ with $\gamma^\prime(0) = X$, we have $\left| \frac{j(\gamma(t))}{t} \right|_g \leq \left| \frac{\rho(\gamma(t))}{t} \right| \overset{t \to 0}{\longrightarrow} 0$.
	In the same fashion, using $|\hat{T}|_g \leq \rho$, we also obtain that $\hat{T}$ is differentiable in $p$ with $\nabla \hat{T}_{|p} = 0$.
	On the other hand, away from the set $\{q \in M| \rho_q=0 \text{ and } \upd_q \rho = 0\}$, where $\nabla \hat{T} = 0$, the $1$-form $\frac{j}{\rho}$ is smooth and
	\begin{gather*}
	\nabla_X \hat{T}
	 	= \nabla_X j \otimes \frac{j}{\rho} + \frac{j}{\rho} \otimes \nabla_X j - \upd \rho(X) \frac{j}{\rho} \otimes\frac{j}{\rho}.
	\end{gather*}
	holds for all $X \in \Gamma(TM)$.
	From this we see that $\nabla \hat{T}_{|q}$ tends to zero for $q \to p$ and so $\nabla \hat{T}$ is continuous in $p$.
\end{Bem}

\begin{Bem}
	If $M$ is $1$-dimensional, a variant of the argument for $\upd_p \rho \neq 0$ also works in the case $\upd_p \rho = 0$, showing that $\hat{T}$ is smooth everywhere.
	We do not spell this out.
\end{Bem}

The central idea for the proof of the main theorem is that certain quotients of smooth functions are not $C^2$, even if the numerator is dominated by the square of the denominator.
For instance, the function $f \colon \R^2 \to \R,\,0 \mapsto 0,\,0 \neq (x,y) \mapsto \frac{x^4}{x^2+2y^2}$ is an example:
Its second partial derivative $\frac{\del^2}{\del x^2}f$ has no continuous extension in zero.
We write down an initial data set $(g,k)$ such that $\rho = |j|_g$ and one of the components of $\hat{T} = \frac{j \otimes j}{\rho}$ is of that kind.

\begin{proof}[Proof of \cref{Thm:Main}]
	For $B_1(0) = \{(x,y) \in \R^2 \mid x^2+y^2 < 1\}$ we consider the following initial data set on $B_1(0) \times \R$:
	The metric $g$ is the Euclidean metric and $k = k_0 + \hat{k}$ for
	\begin{align*}
		k_0(x,y,z) &=
		\begin{pmatrix}
		0 & x^2y & 2xy^2\\
		x^2y & 0 & 0 \\
		2xy^2 & 0 & 0 
		\end{pmatrix}, &
		\hat{k}(x,y,z) &=
		\begin{pmatrix}
		G + c & 0 & 0\\
		0 & G + c & 0 \\
		0 & 0 & -G 
		\end{pmatrix}
	\end{align*}
	with $c \in \R$ and $G = \sqrt{c^2 - x^2 - 2y^2 - x^4y^2 - 4x^2y^4}$, which is smooth for $x^2 + y^2 < 1$ and $c^2 \geq 5$.
	This definition is made in such a way that
	\begin{align*}
		2\rho(x,y,z) &= (G+2c)^2 - (2(G+c)^2 + G^2 + 2(x^2y)^2 + 2(2xy^2)^2) \\
			&= -2G^2 + 2c^2 -2x^4y^2 - 8x^2y^4 = 2x^2 + 4y^2
	\intertext{while}
		j(x,y,z) &=
		\begin{pmatrix}
			\del_x G \\
			\del_y G \\
			0
		\end{pmatrix} -
		\begin{pmatrix}
			\del_x G + x^2 \\
			2xy + \del_y G \\
			2y^2
		\end{pmatrix}
		= \begin{pmatrix}
		x^2 \\
		2xy \\
		2y^2
		\end{pmatrix}.
	\end{align*}
	Thus $\rho = |j|_g$ as required and $\hat{T}$ from \eqref{eq:DefT} is not $C^2$ since
	\begin{gather*}
		\hat{T}_{xx}(x,y,z) = \begin{cases} \frac{x^4}{x^2+2y^2} &(x,y) \neq 0 \\
									0 &(x,y)=0 \end{cases}
	\end{gather*}
	is not.
	This example generalizes to $B_1(0) \times \R^{n-2}$ for $n \geq 3$ by adding zeros to $k_0$ and $\hat{k}$.
	
	In order to transplant it to an arbitrary manifold, we adjust this example a bit further.
	Let $\tilde{\chi}_1$ and $\tilde{\chi}_2$ be smooth cut-off functions $[0, 1) \to [0,1]$ with $\tilde{\chi}_1 \equiv 1$ on $[0,\frac{1}{4}]$, $\tilde{\chi}_1 \equiv 0$ on $[\frac{1}{2}, 1)$, $\tilde{\chi}_2 \equiv 1$ on $[0,\frac{1}{2}]$ and $\tilde{\chi}_2 \equiv 0$ on $[\frac{3}{4}, 1)$.
	Again, we let $g$ be the Euclidean metric on $B_1(0) \times \R^{n-2}$ and set
	\begin{gather*}
		k = \chi_2 k_0 + \chi_1 \hat{k} + (1 -\chi_1) \frac{c}{n-1}g
	\end{gather*}
	for $\chi_i \coloneqq \tilde{\chi}_i(x^2+y^2)$, $i=1,2$.
	Note that on $U^\prime \times \R^{n-2}$ for $U^\prime = \{(x,y) \in \R^2 \mid x^2+y^2 < \frac{1}{4}\}$ we retain the example from above and so the desired properties are satisfied there.
	
	To verify the dec for $\frac{1}{4} \leq x^2 + y^2 \leq \frac{1}{2}$, we first observe that $j$ is left unchanged there since the difference term $(1-\chi_1)(\hat{k}-\frac{c}{n-1}g)$ is a diagonal matrix, only depends on $x$ and $y$, and its trace $(1-\chi_1)(G+\frac{n-2}{n-1}c)$ coincides with its first two entries.
	Thus the additional contribution vanishes in the same way as the $G$-dependent terms canceled the in the calculation of $j$ above.
	For any $p=(x,y,z_3, \ldots, z_n) \in \R^n$ with $(x,y)$ in the range above, let us look at
	\begin{gather*}
		R(t) = \frac{1}{2}\left(t\tr(\hat{k}_p) + (1-t)\frac{n}{n-1}c \right)^2 - \frac{1}{2}|(k_0)_p|_g^2 - \frac{1}{2}\left| t\hat{k}_p +(1-t)\frac{c}{n-1}g_p\right|_g^2
	\end{gather*}
	for $t \in [0,1]$.
	Notice that $\rho(p) = R(\tilde{\chi}_1(x^2+y^2))$.
	We already know that $R(1) = |j_p|_g$.
	For sufficiently large $c$ also $R(0) = \frac{n}{2(n-1)}c^2 - \frac{1}{2}|(k_0)_p|_g^2 \geq |j_p|_g$ holds.
	Since
	\begin{align*}
		\frac{\upd^2}{\upd t^2} R(t) &= \left(G(p)+\frac{n-2}{n-1}c\right)^2 \\&\phantom{=}- 2\left(G(p) +\frac{n-2}{n-1}c\right)^2 - \left(-G(p) -\frac{c}{n-1}\right)^2 -(n-3)\left(\frac{c}{n-1}\right)^2 
	\end{align*}
	is negative, $R$ is concave in $t$ and thus $R(t) \geq |j_p|_g$ holds for all $t \in [0,1]$.
	In particular, dec is satisfied for $\frac{1}{4} \leq x^2+y^2 \leq \frac{1}{2}$.

	On the section $\frac{1}{2} \leq x^2+y^2 \leq \frac{3}{4}$, both $\chi_2^2|k_0|_g^2$ and $|j|_g$ are bounded since they are continuous functions (just) depending on $x$ and $y$ and these range over a compact set.
	Due to $\rho = \frac{n}{2(n-1)}c^2 - \frac{1}{2}\chi_2^2 |k_0|_g^2$, dec is satisfied there if $c$ is chosen suitably large.
	Finally, for $\frac{3}{4} \leq x^2+y^2 < 1$ we have $\rho = \frac{n}{2(n-1)}c^2$ and $j = 0$, so dec is evident.
	  
	Now let $M$ be a manifold of dimension $n \geq 3$ and choose a Riemannian metric $g$ on $M$ with bounded curvature.
	We may moreover assume that there is a subset $V$ diffeomorphic to $B_1(0) \times (\R^{n-2}/\Z^{n-2})$ where the pulled-back metric is the Euclidean one.
	In fact, this thickened torus could just lie within a small ball.
	In the following, we will identify $B_1(0) \times (\R^{n-2}/\Z^{n-2})$ with $V$.
	We note that the symmetric $2$-tensor $k$ defined above descends to $B_1(0) \times (\R^{n-2}/\Z^{n-2})$ and extends smoothly to $M$ setting $k = \frac{1}{n-1} cg$ on $M \setminus V$.
	By what has been established above, the initial data set $(g,k)$ on $M$ has the demanded properties on $U = U^\prime \times (\R^{n-2}/\Z^{n-2}) \subset V$ and satisfies dec on $V$.
	Since the curvature of $(M,g)$ is bounded, $\rho = \frac{1}{2}\scal^g + \frac{n}{2(n-1)}c^2$ and $j = 0$ on $M \setminus V$, dec is also satisfied on the rest of $M$ provided $c$ is large enough.
	
	For the final claim note that if $(\overline{M}, \overline{g})$ is a smooth dec spacetime and $M \subset \overline{M}$ is an embedded spacelike hypersurface with induced initial data set $(g,k)$, then $\hat{T} \coloneqq \ein_{|TM \otimes TM}$ satisfies \eqref{eq:dec}.
	Thus if $\rho = |j|_g$ on some open subset $U$, then $\hat{T}_{|U}$ coincides with the expression from \eqref{eq:DefT} according to \cref{Lem:Rigid}.
	This yields the desired contradiction if this expression does not define a smooth section on $U \neq \emptyset$.
\end{proof}

\begin{Bem}
As a byproduct, the initial data set constructed in the proof of \cref{Thm:Main} features $\tr(k) \geq \frac{n}{n-1}c > 0$ and thus provides initial conditions for a big bang type singularity according to Hawking's singularity theorem \cite[Sec.~8.2, Thm.~4, condition (3')]{Hawking.Ellis:1973}.
(Clearly, the sign of $k$ can also easily be reversed.)
However, since the gluing in the proof happens within a small open neighborhood, it should be possible to construct initial data sets as in \cref{Thm:Main} with other desired properties.
For example, one could try and prescribe the asymptotic behavior of $(g,k)$.
When doing so, of course, the restrictions imposed by the positive mass theorem should be respected.
\end{Bem}

\end{document}